\documentclass[a4paper, conference]{IEEEtran}
\usepackage{cite}
\usepackage{amsmath}
\usepackage{caption2}
\usepackage{subfigure}

\usepackage{geometry}
\makeatletter
\def\fnum@figure#1{\figurename\nobreakspace\thefigure\hspace{1em}}
\def\fnum@table#1{\tablename\nobreakspace\thetable\hspace{1em}}
\makeatother
\ifCLASSINFOpdf
  \usepackage[pdftex]{graphicx}
\else
\fi
\begin{document}
%
\title{An SVD-based Fragile Watermarking Scheme With Grouped Blocks}

\author{\IEEEauthorblockN{Qingbo Kang}
\IEEEauthorblockA{Chengdu Yufei Information\\Engineering Co.,Ltd.\\
610000 \ Chengdu, China\\
Email: qdsclove@gmail.com}
\and
\IEEEauthorblockN{Ke Li, Hu Chen}
\IEEEauthorblockA{ National Key Laboratory of Fundamental \\ Science on Synthetic Vision.\\
Sichuan University, 610000 \ Chengdu, China\\ Email: likeneill@gmail.com, huchen@scu.edu.cn} }


%


\maketitle

\begin{abstract}
This paper proposes a novel fragile watermarking scheme for digital image authentication which is based on Singular Value Decomposition(SVD) and grouped blocks. The watermark bits which include two types of bits are inserted into the least significant bit(LSB) plane of the host image using the adaptive chaotic map to determine the positions. The groped blocks break the block-wise independence and therefore can withstand the Vector Quantization attack(VQ attack). The inserting positions are related to the statistical information of image block data, in order to increase the security and provide an auxiliary way to authenticate the image data. The effectiveness of the proposed scheme is checked by a variety of attacks, and the experimental results prove that it has a remarkable tamper detection ability and also has a precise locating ability.
\end{abstract}

\begin{IEEEkeywords} Image Authentication, Tamper Detection, Fragile Watermarking, Singular Value Decomposition
\end{IEEEkeywords}

%
\IEEEpeerreviewmaketitle

\section{Introduction}
With the tremendous development of information technology, especially in network communication and multimedia, digital images have a paramount role in our daily life. However, the digital images can be easily modified and tampered with the help of powerful image processing software. In fact, lots of people can easily manipulate images that may bring about human casualty or financial loss\cite{haouzia2008methods}. So maintaining the authenticity and integrity of digital images has become a considerable aspect of many organizations \cite{lin2011multimedia, friedman1993trustworthy}. 
The image authentication schemes can fall into two types according to the methods they are based on: cryptography based schemes\cite{matsuo2004parallel, li2005oblivious} and fragile watermark based schemes\cite{yeung1997invisible, wong2001secret, suthaharan2004fragile}. Image authentication schemes based on cryptography calculate a message authentication code (MAC) from images using a hash function, and they can detect if an image has been modified, but they don't have the ability to locate the modified regions\cite{lu2003structural}. 
In a scheme which is based on fragile watermark, the host images that need to be protected are embedded with the watermark which is often generated using either image features extracted from host image or the generated random values, when the image need to be authenticated, the original watermark is then extracted from the watermarked image to detect the tampered regions\cite{hu2013probability}. Walton proposed the first fragile watermark-based authentication schemes\cite{2023640}. It only provides very limited tamper detection\cite{liu2007image}. Holliman and Memon proved that schemes which are block-wise independent are vulnerable to vector quantization attack\cite{holliman2000counterfeiting}. The Vector Quantization attack(VQ attack) means the counterfeit image can be reconstructed using a vector quantization code-book generated from a set of watermarked images, because all blocks in the images are authenticated, hence the counterfeit images are authentic with the watermarking scheme. To withstand the VQ attack, researches proposed a number of schemes. \cite{liu2007image, 49593321} proposed the fragile watermarking schemes that use the chaotic pattern to generate the different image and then map it into a binary image eventually insert into the LSB bit-plane of the host image. Since the corresponding watermark of the modified pixel value may be consistent with the original watermark, result in the failure detection to the tampered image. In this case, localization of the tampered regions is incomplete, and detection of the tampering pattern is imprecision\cite{5813456}.

Singular value decomposition(SVD) is a kind of effective method of algebraic feature extraction. It can not only capture the basic structure of the data in the matrix, but also reflect the algebraic essence of the matrix. These excellent features make it have a wide application in signal processing, image compression, pattern recognition and other fields. SVD also has been widely used in robust watermarking field\cite{lai2011digital}. In the scheme proposed by Sun et.al\cite{sun2002svd}, performing SVD in the spatial domain, and the watermark is embedded by quantizing the largest SV of an image block. However, this method is vulnerable to VQ attack. In this paper, a novel SVD-based watermarking scheme for image authentication is proposed. The blocks of the host image are disturbed with the help of Arnold scrambling method. Then, all scrambled image blocks are divided into grouped blocks. For each block, two types of watermark bits are embedded, one for the block itself, the other for the grouped blocks. An adaptive chaotic image pattern is generated using the logistic map for each block to determining the embedded position of the watermark bits. The use of the watermark bits of the grouped blocks is to break the block-wise independence and withstand the VQ attack.


\section{Singular Value Decomposition and Chaotic Maps}

\subsection{Singular Value Decomposition}
In linear algebra, the SVD is a factorization of a real or complex matrix. Formally, any real or complex $m \times n$ matrix $M$ of rank $r$ can be decomposed as
\begin{equation}\label{eq_svd}
M = USV^{T}
\end{equation}
where $U_{m \times m}$ and $V_{n \times n}$ are unitary matrices, and $S_{m \times n}$ is an $m \times n$ rectangular diagonal matrix. Moreover, 
\begin{displaymath}
S_{m \times n} = \begin{bmatrix}
                 \triangle_{r \times r} & 0 \\
                 0 & 0
                 \end{bmatrix} 
\end{displaymath}
\begin{equation}
\triangle_{r \times r} = diag(\sigma_{1}, \sigma_{2}, \ldots, \sigma_{r})
\end{equation}
\begin{displaymath}
\sigma_{i} = \sqrt{\lambda_{i}}(i = 1, 2, \ldots, r, \ldots, n)
\end{displaymath}

The diagonal entries $\sigma_{i}$ of $\triangle_{r \times r}$  are known as the singular value of the matrix $M$. $\lambda_{1} \geq \lambda_{2} \geq \ldots \geq \lambda_{r} \geq 0, \lambda_{r + 1} = \lambda_{r + 2} = \ldots = \lambda_{n} = 0$ are the eigenvalues of both $M^{T}M$ and $MM^{T}$. Under the limitation of $\lambda_{1} \geq \lambda_{2} \geq \ldots \geq \lambda_{r}$, the vector $(\sigma_{1}, \sigma_{2}, \ldots, \sigma_{r})$ is unique. It characterizes the distribution and retains the algebraic essence of the matrix data\cite{1646545}.
\subsection{Chaotic Maps}
In recent years, chaotic system and permutation transform have been commonly used in digital watermarking in order to enhance the security\cite{49593321}. Use the logistic map and Arnold scrambling to increase the security and performance of our scheme.
\subsubsection{Logistic Map}
Logistic map is one of the simplest and most transparent systems exhibiting order to chaos transition. Mathematically it is defined as:
\begin{equation}\label{eq1}
x_{n+1} = \mu x_{n}(1 - x_{n}), n \in Z, u \in [0, 4], x_{n} \in (0, 1)
\end{equation}
The $\mu$ here is a positive constant sometimes known as the ``biotic potential'', when $3.5699456 < \mu < 4$ the map is in the region of fully developed chaos\cite{18569710}. That is, at this point, the sequence ${x_{k}; k = 0, 1, 2, 3, \ldots}$ generated by \eqref{eq1} is non-periodic, non-convergent and sensitive to the initial value.
\subsubsection{Arnold Transform} \label{Arnold}
Arnold transform has been widely used in the field of image encryption. The classical Arnold transform is a two-dimensional invertible chaotic map described by
\begin{equation}\label{eq2}
\begin{bmatrix}x_{n}\\ y_{n} \end{bmatrix} = \begin{bmatrix}1 & a \\ b & ab+1 \end{bmatrix} \begin{bmatrix}x_{0}\\ y_{0} \end{bmatrix} (mod\ N)
\end{equation}
where $a$ and $b$ are positive integers, $N$ represents the size of the image matrix, it indicates the start point $(x_{0}, y_{0})$ after $n$ iterations, then through modular arithmetic to get the coordinates of the final result. The \eqref{eq2} is also chaotic and area preserving. The parameters $a, b$ and the iterations $k$, can serve as secret keys to enhance the security, too.

\section{The Fragile Watermarking Scheme}
In the proposed scheme, the watermark bits to be embedded in the carrier image consists of two parts: block authentication bits for authenticating the image block itself, and group authentication bits for authenticating the grouped blocks. Our scheme is a block-based scheme. In order to withstand the VQ attack, the grouped blocks is used to break the independence between the blocks. For each one image block, these two types of authentication bits are hidden in the LSB of the block pixels. The positions used for insertion are determined by the chaotic sequence, which depends on the statistical information of the block pixels. If the statistical information is changed, all the watermark bits can't be correctly extracted from the image block. This is an auxiliary way to authenticate the image data. So the use of chaotic sequence not only increases the security, but also increases the creditability of the proposed scheme.

\begin{figure*}
\centering
\includegraphics[]{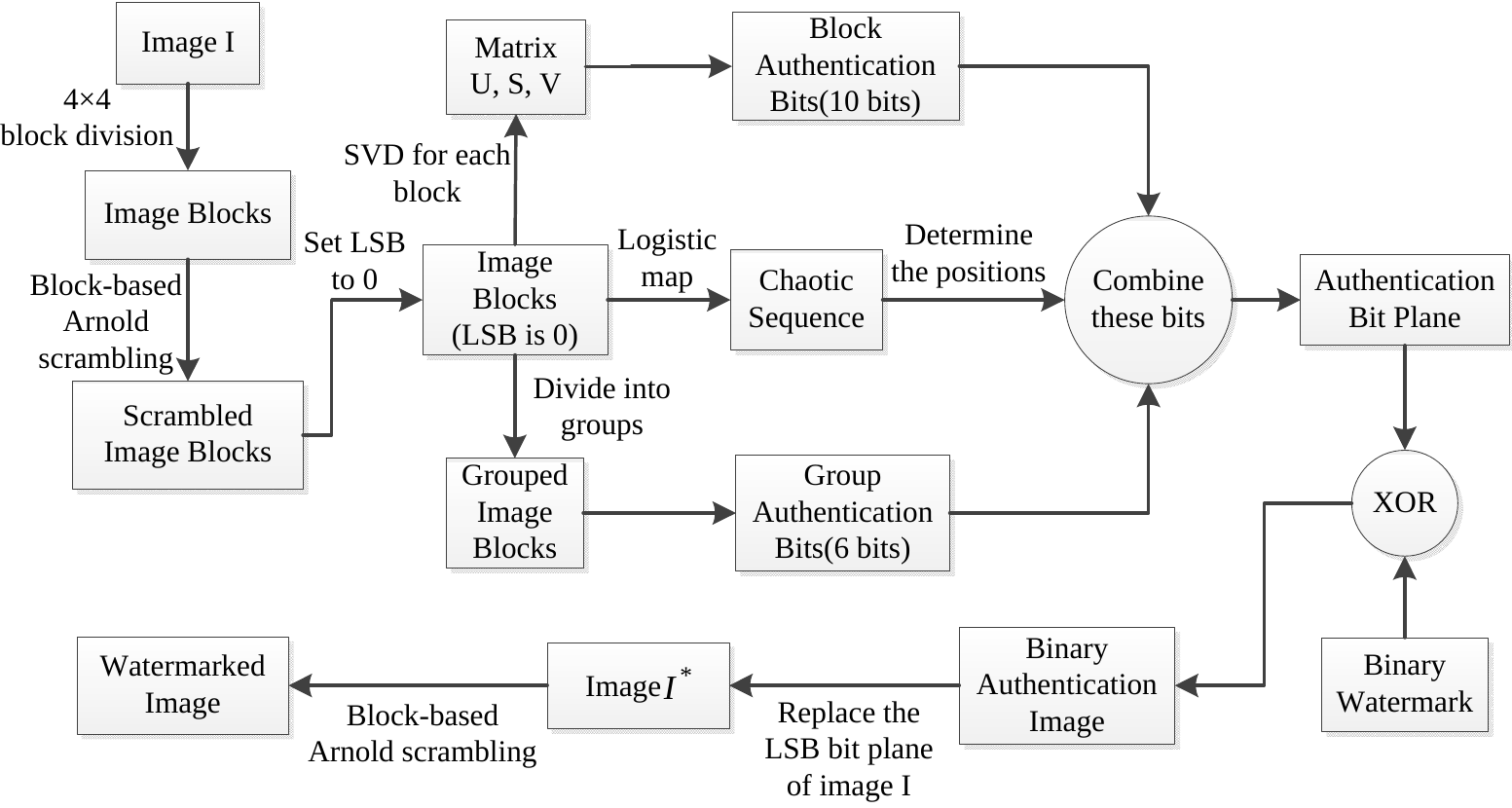}

\caption{Block diagram of embedding procedure}
\label{fig_embed}
\end{figure*}

Figure \ref{fig_embed} shows the block diagram of the embedding procedure. The more details are described in below.

\subsection{Watermark Embedding Procedure}
\subsubsection{Block Division}
Before generating and embedding watermark bits, first the original image has to be divided into blocks. Denote the original grayscale image as $I$, and its rows and columns are $M_{1}$ and $M_{2}$,
so $M(M = M_{1} \times M_{2})$ is the total number of image pixels. Assuming that both $M_{1}$ and $M_{2}$ are the whole multiple of four, first divide the image into $M/16$ non-overlapped blocks, one block is sized $4 \times 4$, and represents these blocks as $B_{m, n}(m \in [1, M_{1}/4], n \in [1, M_{2}/4] )$ and the pixel values in one block as $b_{m, n}(i, j)(1 \leq i, j \leq 4)$. 

\subsubsection{Arnold Scrambling}
This step the Arnold transform described in \eqref{eq2} is used to scramble the original image blocks $B_{m, n}$. The unit of scrambling is image block. The times of transformation is $k$, where $k$ is the security number. Let's denote the scrambled image as $ScrI$ and the blocks in the scrambled image as $ScrB_{m, n}(m \in [1, M_{1}/4], n \in [1, M_{2}/4])$, and the values of pixels in a block as $Scrb_{m, n}(i, j)(1 \leq i, j \leq 4)$.

\subsubsection{Set LSB to Zero}
For all image blocks in  $ScrB_{m, n}$, Set LSB bitplane of pixels as zero: 
\begin{equation}
\begin{aligned}
Scrb_{m, n}(i, i) = Scrb_{m, n}(i, j) - (Scrb_{m, n}(i, j)\ mod\ 2). \\ 
(m \in [1, M_{1}/4], n \in [1, M_{2}/4]), (1 \leq i, j \leq 4).
\end{aligned}
\end{equation}

\subsubsection{SVD on The Image Blocks}
Treat the image pixels in one block as a matrix, and perform SVD on the image blocks. We obtain three matrixes $U, S, V$ as in \eqref{eq_svd} for each one block. Then, the trace of the matrix S is calculated, eventually we map the traces to the range $[0, 1023]$ and we name the value as Block Authentication Number(BAN). Assuming the traces are $Trace_{m, n}(m \in [1, M_{1}/4], n \in [1, M_{2}/4])$, i.e.,
\begin{equation}\label{BAN}
BAN_{m, n} = \lfloor(Trace_{m, n} mod \ 1024)\rfloor
\end{equation}
\subsubsection{Group The Image Blocks}
Grouping all the scrambled image blocks $ScrB_{m, n}$, each group have five blocks. The detailed procedure of Grouping is described as follows.

Firstly, the index of scrambled image block is converted from two-dimensional to one-dimensional, i.e.,
\begin{displaymath}
\begin{aligned}
ScrB_{k} = Scrb_{m, n}
\end{aligned}
\end{displaymath}
\begin{equation}
k = (m - 1) \times M_{1} / 4 + n
\end{equation}
\begin{displaymath}
(m \in [1, M_{1}/4],n \in [1, M_{2}/4]), k \in [1, M]
\end{displaymath}
Then, for each block in $ScrB_{k}$, we calculate the start position $StartIndex$ and end position $EndIndex$ of the group in which the block is:
\begin{equation}
StartIndex_{k} = \lfloor ((k - 1) / 5)\rfloor \times 5 + 1
\end{equation}
\begin{equation}
EndIndex_{k} = \lceil (k / 5)\rceil \times 5, \ k \in [1, M]
\end{equation}
Thirdly, for every block, we obtain the grouped scrambled image blocks $GSB_{k}$:
\begin{equation}
GSB_{k} = {ScrB_{GroupIndexes}}
\end{equation}
\begin{equation}
GroupIndexes \in [StartIndex_{k}, EndIndex_{k}]
\end{equation}
Finally, the index of grouped scrambled image blocks is converted from one-dimension to two-dimension, i.e.,
\begin{displaymath}
\begin{aligned}
GSB_{p, q} = GSB_{k}
\end{aligned}
\end{displaymath}
\begin{equation}
p = \lceil k / (M_{2}/4) \rceil, q = k \ mod \ (M_{2}/4)
\end{equation}
\begin{displaymath}
p \in [1, M_{1}/4],\ q \in [1, M_{2}/4],\ k \in [1, M]
\end{displaymath}

\begin{figure}
\centering
\subfigure[scrambled image]{\includegraphics[width=0.22\textwidth]{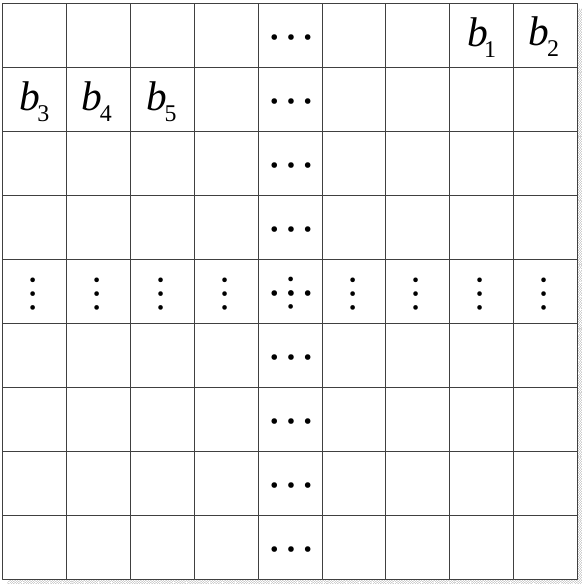}} \quad
\subfigure[original image]{\includegraphics[width=0.22\textwidth]{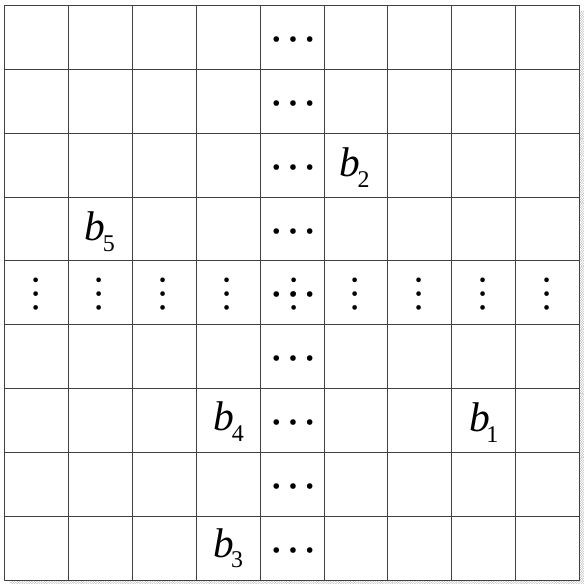}}
\caption{Grouped blocks in scrambled image and original image}
\label{fig_grouped}
\end{figure}

Figure \ref{fig_grouped} illustrates the basic concept of image blocks grouping method. In Figure \ref{fig_grouped}, $\{b_{1}, b_{2}, b_{3}, b_{4}, b_{5}\}$ are grouped image blocks. They are neighbours in the scrambled image, but they are spatially unrelated in the original image. Since the scrambling times is not known to the attacker, hence he can't obtain the grouped image blocks without the key. It improves the security of our scheme.
\begin{figure*}
\centering
\includegraphics[width = 1.0\textwidth]{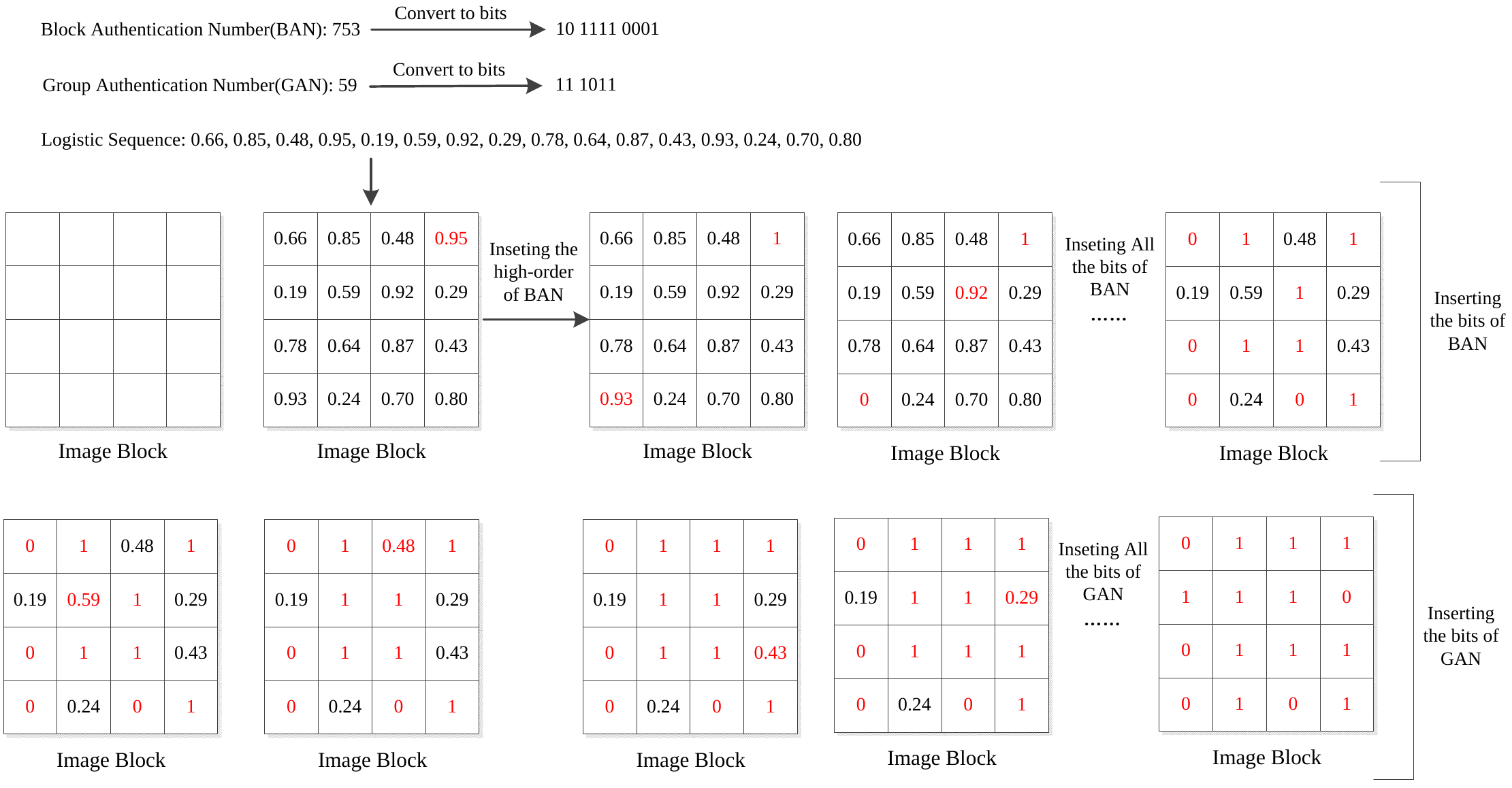}

\caption{An example of watermark bits inserting}
\label{fig_insert}
\end{figure*}
\subsubsection{Calculate Group Authentication Number}
According to the Grouped image blocks $GSB_{p, q}$ and Block authentication number $BAN_{m, n}$ that we have got, we can calculate the Group Authentication Number(GAN) for each group, that is:
\begin{equation}
GAN_{p, q} = \sum_{(m, n)\ is\ in\ GSB_{p, q}} BAN_{m, n} / 5
\end{equation}
Eventually we map the GAN to the range $[0, 63]$:
\begin{equation}
GAN_{p, q} = \lfloor GAN_{p, q}\ mod\ 64 \rfloor , 
\end{equation}
\begin{equation}
p \in [1, M_{1}/4],\ q \in [1, M_{2}/4]
\end{equation}

\subsubsection{Generate The Adaptive Chaotic Sequence}
For one block in the $ScrB_{m, n}$, compute the average value and the standard deviation of the pixels. Denote them as $Average_{m, n}$ and $StDev_{m, n}$, respectively. Then:
\begin{equation}
Initial_{m, n} = (Average_{m, n} + 1)/257
\end{equation}
\begin{equation}
Param_{m, n} = 3.5699 + (StDev_{m, n} - \lfloor StDev_{m, n} \rfloor) \times 0.43
\end{equation}
In this way we can obtain the initial value and $\mu$ of logistic sequence described in \eqref{eq1}. At last we use these above values to generate the logistic sequences $LogSeq_{m, n}$, wherein the length of any one is 16:
\begin{equation}
LogSeq_{m, n} = \{ x_{1}, x_{2}, \ldots, \ x_{16} \}
\end{equation}
This mechanism ensures that the different image blocks which have different average value or standard deviation obtain different logistic sequence.

\subsubsection{Insert Watermark}
First convert the BAN and GAN to binary bits. Since BAN are in the range [0, 1023] and GAN are in the range [0, 63], in that way the total length of these two binary bits is 16. The embedding positions depend on the logistic sequence, where the maximum value of the logistic sequence inserted with the high-order bit of BAN, and so forth. After inserting all bits of BAN, then the bits of GAN are inserted one by one depend on the relationship of the logistic sequence. The low-order bit of GAN insert to the position where the minimum value of the logistic sequence is. Figure \ref{fig_insert} give an example of inserting the watermark bits in one image block. 

After all blocks are done watermark inserting, merge all $4 \times 4$ bit planes to obtain the Authentication Bit Plane $(ABP)$, which has the same with the original image.

\subsubsection{XOR with the Binary Watermark Image}
The watermark image is a visually meaningful binary image and denote it as $W$. The size of $W$ is the same with the original image $I$. Obtain the binary authentication image $BAI$ using XOR operation between the binary image $W$ and $ABP$ as follows:
\begin{equation}
BAI = W \oplus ABP
\end{equation}
\subsubsection{  Replace the LSB}
Replace the LSB plane of $ScrI$ by $BAI$.
\begin{figure*}
\centering
\includegraphics[width=0.95\textwidth]{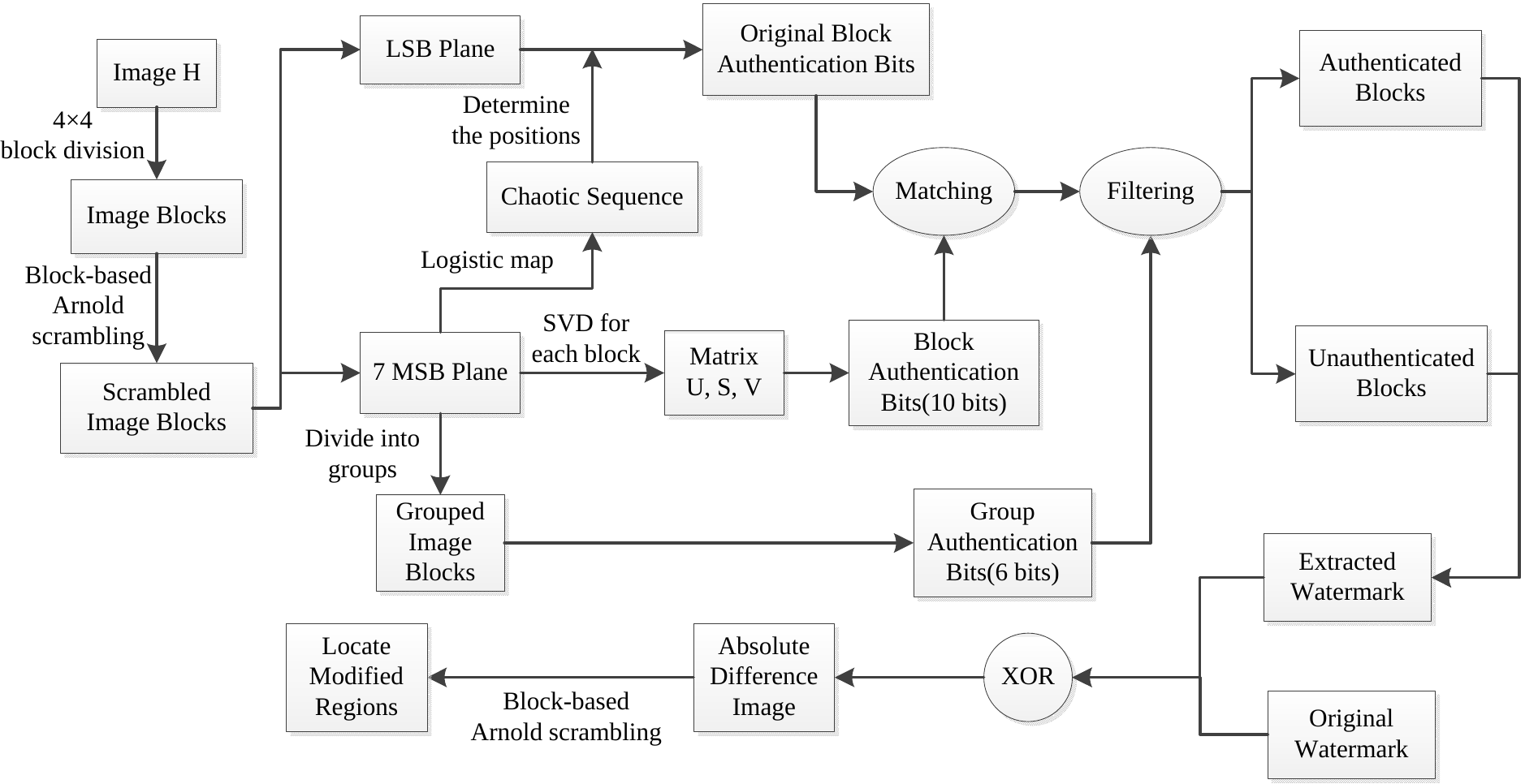}
\caption{Block diagram of extracting procedure}
\label{fig_extract}
\end{figure*}
\subsubsection{  Arnold Scrambling}
Owing to the Arnold transform is a periodic transformation, $(T - k)$ times of it is employed on $ScrI$ to obtain the watermarked image, where $T$ is the period of the Arnold transform.

\subsection{Watermark Extracting Procedure}
Figure \ref{fig_extract} schematically shows the watermark extract procedure. Most of the steps are same or similar to the corresponding steps of the embedding procedure, therefore these steps will not repeat description here. 

With the LSB plane of watermarked image, combined with the chaotic sequences generated by block to determining to positions, the original block authentication bits for each image block is obtained and denote it as $OrigBAN_{m, n}$. The block authentication bits and the group authentication bits calculated from the 7 MSB plane are $BAN_{m, n}$ and $GAN_{m, n},  (m \in [1, M_{1}/4],n \in [1, M_{2}/4])$, respectively. The procedure of matching to obtain the extracted watermark image and locate the modified regions of watermarked image is described as follows.
\subsubsection{Block Authentication Bits Matching}
This step is very simple and straightforward. If the original block authentication bits is not equal to the calculated block authentication bits, then the block is an unauthenticated block. Meanwhile the watermark block generated by the block is in contrast with the original watermark at the same position. The extracted watermark $EW_{m, n}$ is a binary image and the original watermark $OW_{m, n}$ also is an binary image.
\begin{equation}
EW_{m, n} = \begin{cases} OW_{m, n} & if \ BAN_{m, n} = OrigBAN_{m, n} \\ \sim OW_{m, n} & otherwise \end{cases}
\end{equation}
\subsubsection{Group Authentication Bits Filtering}
Firstly, divide all the blocks into groups using the method described above. Find the most frequent element in one group, and denote it as $FreqGAN_{m, n}$. Then, 
\begin{equation}
EW_{m, n} = \begin{cases} OW_{m, n} & if \ GAN_{m, n} = FreqGAN_{m, n} \\ \sim OW_{m, n} & otherwise \end{cases}
\end{equation}
\subsubsection{Locating the Modified Regions}
Merge the $EW_{m, n}$ to get the complete extracted watermark binary image $W_{ext}$, next apply XOR operation to $W_{ext}$ and the original watermark image. Employ Block-based Arnold scrambling $(T - k)$ times to obtain the image of tampered regions.

\section{Simulation Experiments}
In order to assess the performance of the proposed scheme, in this section a variety of attacks are applied to obtain the simulation experiments. The watermark image in the experiments is a $512 \times 512$ binary image, and the parameters of Arnold transform are $a = 1, b = 1$, and $k = 30$. The degree of distortion between the watermarked image and the original image is measured by PSNR(peak signal-to-noise ratio).

\subsection{Copy and paste attack}
\begin{figure*}
\centering
\subfigure[Original image]{\includegraphics[width=0.30\textwidth]{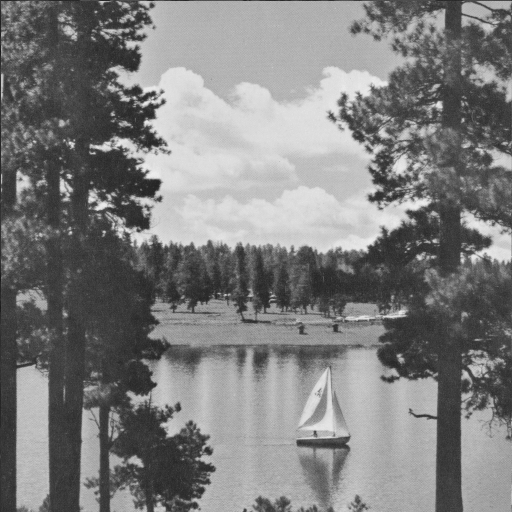}} \quad
\subfigure[Binary watermark image]{\includegraphics[width=0.30\textwidth]{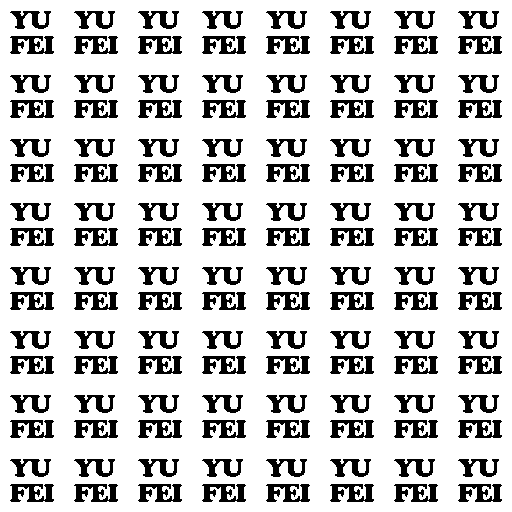}} \quad
\subfigure[Watermarked image]{\includegraphics[width=0.30\textwidth]{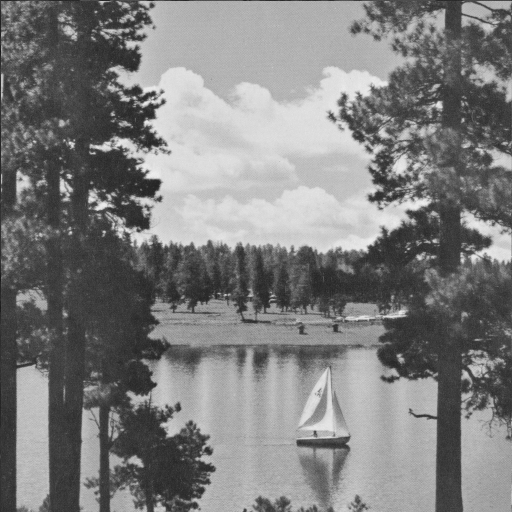}} 
\caption{Watermark embedding experiment}
\label{fig_img1}
\end{figure*}
\begin{figure*}
\centering
\subfigure[Tampered  image]{\includegraphics[width=0.30\textwidth]{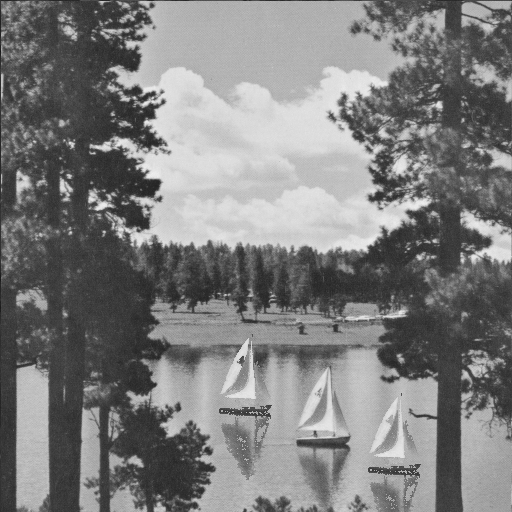}} \quad
\subfigure[Extracted watermark]{\includegraphics[width=0.30\textwidth]{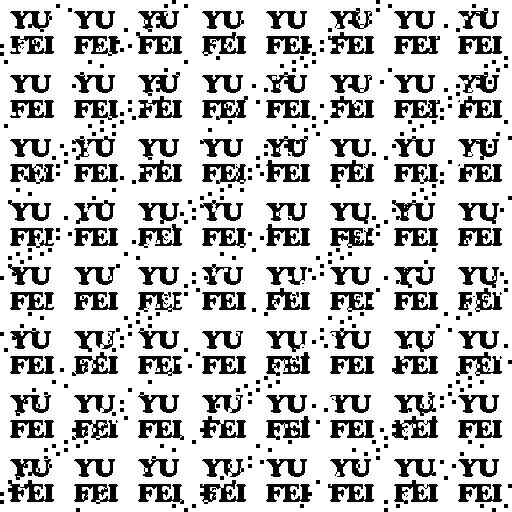}} \quad
\subfigure[Detected tampered region]{\includegraphics[width=0.30\textwidth]{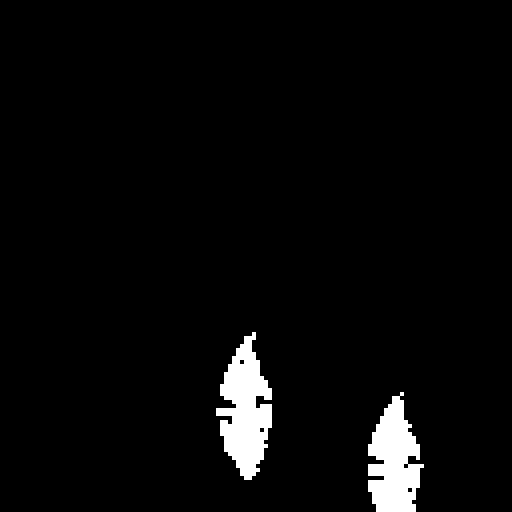}} 
\caption{Watermark extracting experiment}
\label{fig_img2}
\end{figure*}

The original image which needs to be protected is a $512 \times 512$ 'Sailboat' image. Figure \ref{fig_img1} shows the result of watermark embedding procedure. The PSNR of the watermarked image with the original image is 51.1420 dB. There are two kinds of copy and paste attacks. The first is adding two extra sailboats into the watermarked image, and the sailboats are directly copied from the watermarked image. The experiment result is shown in Figure \ref{fig_img2}.

\begin{figure*}
\centering
\subfigure[Tampered  image]{\includegraphics[width=0.30\textwidth]{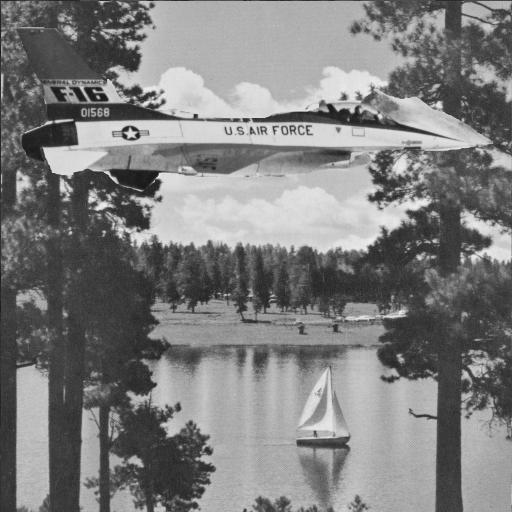}} \quad
\subfigure[Extracted watermark]{\includegraphics[width=0.30\textwidth]{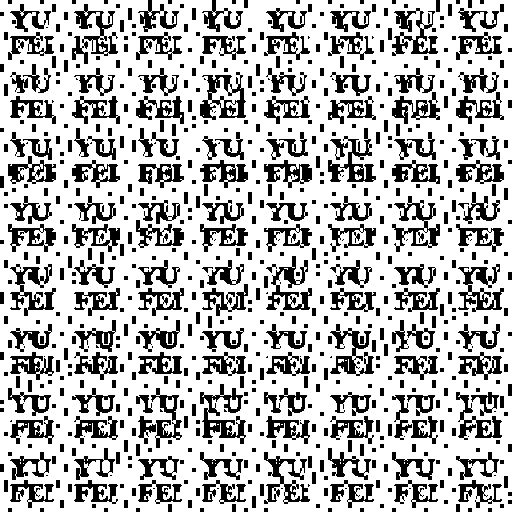}} \quad
\subfigure[Detected tampered region]{\includegraphics[width=0.30\textwidth]{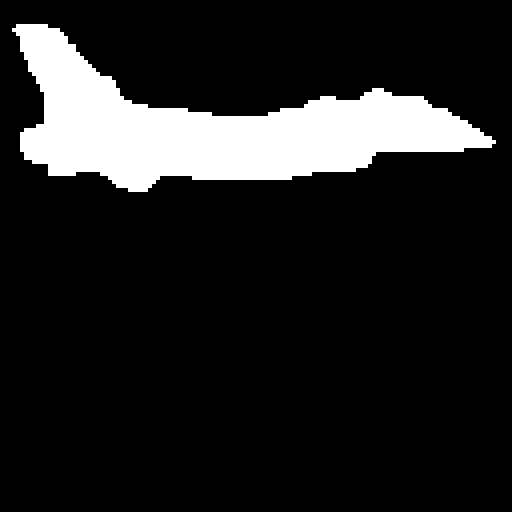}} 
\caption{Watermark extracting experiment}
\label{fig_img3}
\end{figure*}

The second is adding an U.S. Air Force jet into the watermarked image, similarly, the image region of the jet is copied from another watermarked image. The experiment result is displayed in Figure \ref{fig_img3}.
\subsection{Text addition attack}
\begin{figure*}
\centering
\subfigure[Watermarked couple image]{\includegraphics[width=0.22\textwidth]{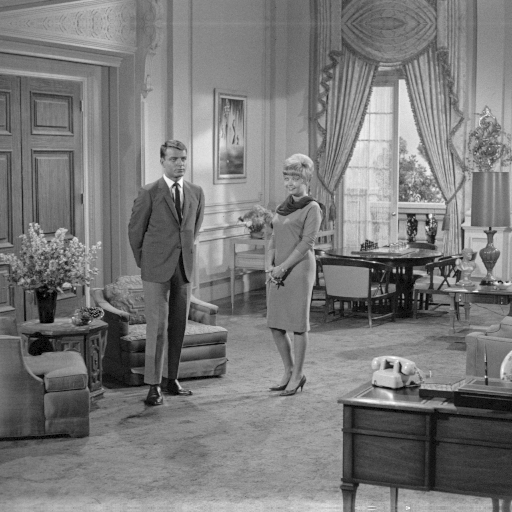}} \quad
\subfigure[Tampered couple image]{\includegraphics[width=0.22\textwidth]{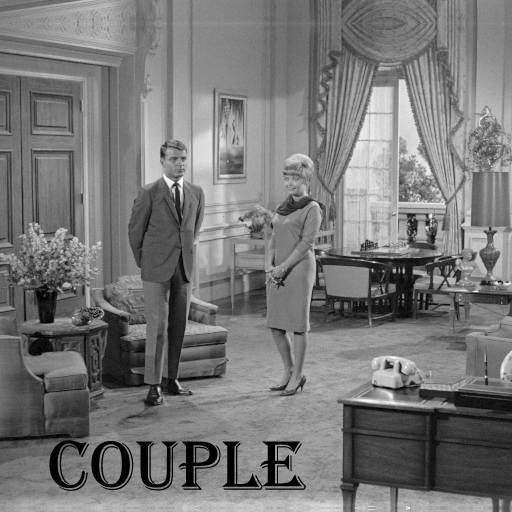}} \quad
\subfigure[Extracted watermark]{\includegraphics[width=0.22\textwidth]{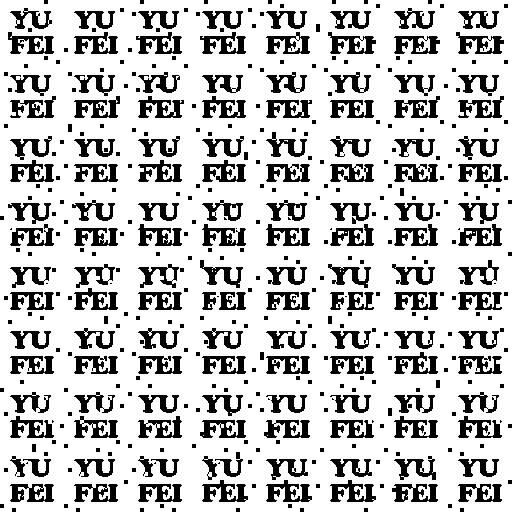}} \quad
\subfigure[Detected tampered region]{\includegraphics[width=0.22\textwidth]{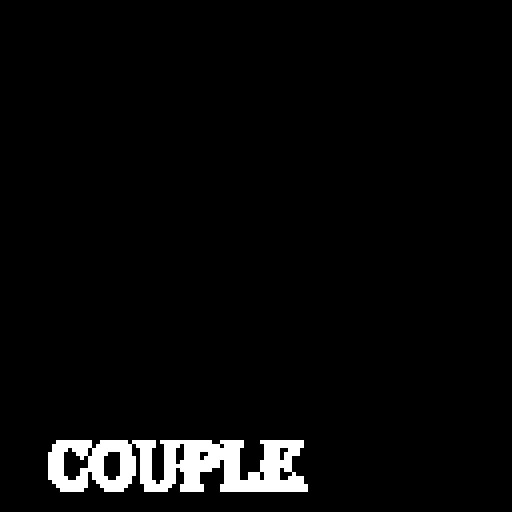}} 
\caption{Experiment result of the text addition attack}
\label{fig_img4}
\end{figure*}
The Figure \ref{fig_img4}(a) is a watermarked image, and the text `COUPLE' is added at the bottom of it as Figure \ref{fig_img4}(b) shows. The extraction result is shown in Figure \ref{fig_img4}(c), and Figure \ref{fig_img4}(d) displayes the tampered region.
\subsection{Content removal attack}
\begin{figure*}
\centering
\subfigure[Tampered couple image]{\includegraphics[width=0.30\textwidth]{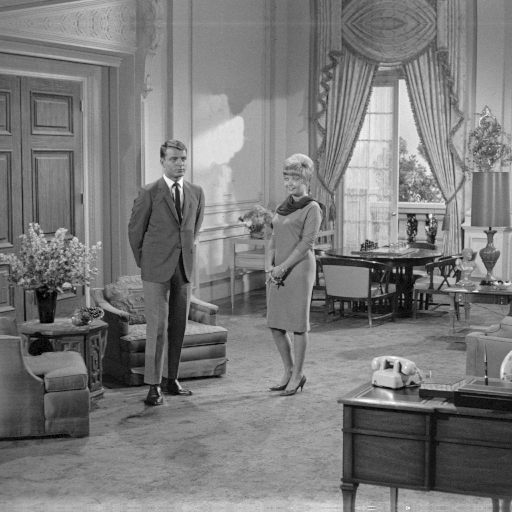}} \quad
\subfigure[Extracted watermark]{\includegraphics[width=0.30\textwidth]{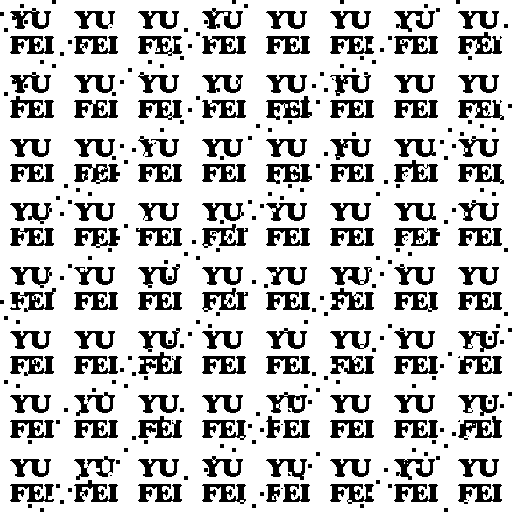}} \quad
\subfigure[Detected tampered region]{\includegraphics[width=0.30\textwidth]{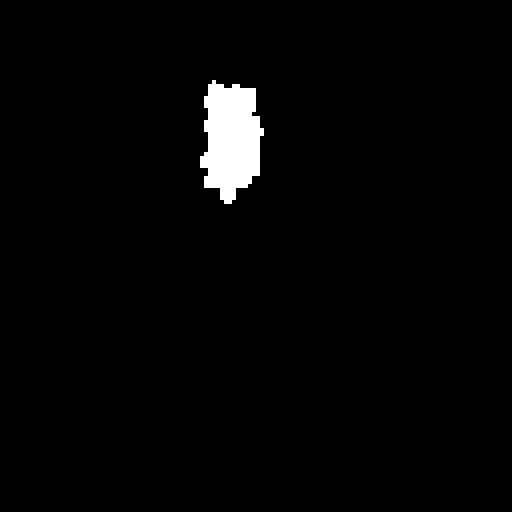}} 
\caption{Experiment result of the content removal attack}
\label{fig_img5}
\end{figure*}

The Figure \ref{fig_img5}(a) is a watermarked image, and the painting hanging on the wall had been removed from the watermarked image. Figure \ref{fig_img5}(b) and Figure \ref{fig_img5}(c) show the extraction result and the tampered region, respectively.
\subsection{VQ attack}
\begin{figure*}
\centering
\subfigure[Tampered couple image with VQ attack]{\includegraphics[width=0.30\textwidth]{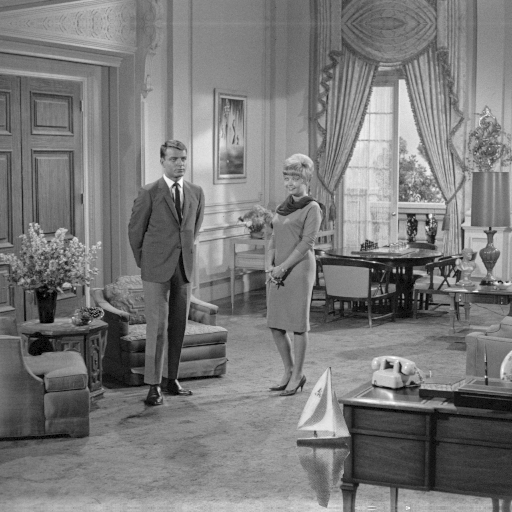}} \quad
\subfigure[Extracted watermark]{\includegraphics[width=0.30\textwidth]{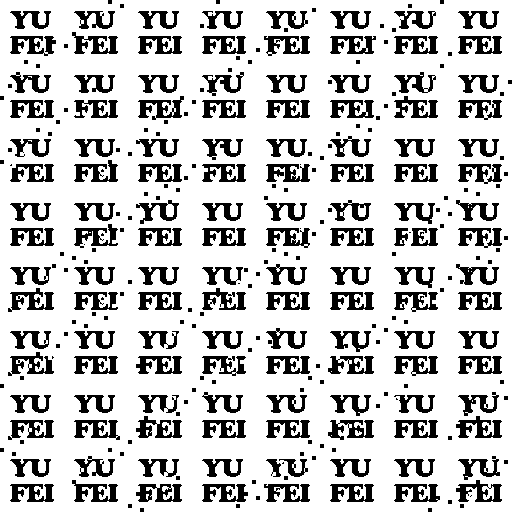}} \quad
\subfigure[Detected tampered region]{\includegraphics[width=0.30\textwidth]{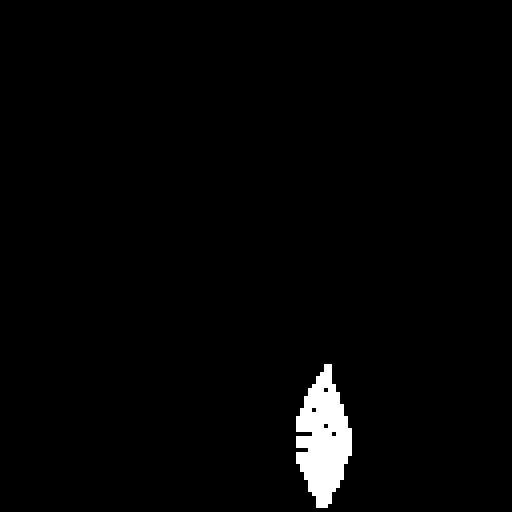}} 
\caption{Experiment result of the VQ attack}
\label{fig_img6}
\end{figure*}

In VQ attack, a forgery is obtained by combining different parts of different watermarked images, while maintaining the position of the portion in the original image \cite{49593321}. Figure \ref{fig_img6} shows the experiment result. The fake image, as shown in Figure \ref{fig_img6}(a) was obtained by copying the sailboat from Figure \ref{fig_img1}(c) and pasting it in Figure \ref{fig_img4}(a).

\section{Conclusion}
In this paper, A novel fragile watermarking scheme based on SVD and grouped blocks for image authentication is proposed. In order to withstand VQ attack, the watermark bits include two types of bits: the block authentication bits for authenticating the block image data and the group authentication bits for authenticating the grouped block image data. The chaotic sequence which is adaptive for each image block is generated in order to increase the security and provides an auxiliary way to authenticate the image block data. The experiment results have shown that the novel scheme, has excellent tamper detection and precise locating capabilities, especially can withstand VQ attack, therefore, has a broad application prospects in the field of image authentication.


\section*{Acknowledgment}
This work was supported by the National Natural Science Foundation of China (Grant No.61202160).



%

\bibliographystyle{unsrt}
\bibliography{references}

\end{document}